\documentclass[intlimits,twoside,a4paper]{article}
\usepackage{graphicx} % Required for inserting images
\usepackage{color}

\usepackage{cmpj3}
\usepackage{multirow}

\issue{2026}{29}{1}{13603}
\doinumber{10.5488/CMP.29.13603}

\title[How does ethane wet different substrates?]{How does ethane wet different substrates?}
\author[\L. Baran, D. Tarasewicz, W. R\.zysko] {\L. Baran\orcid{0000-0003-1777-1998} \refaddr{label1,label2}\thanks{Corresponding author: \email{lukasz.baran@mail.umcs.pl}.},
D. Tarasewicz\orcid{0000-0002-4010-9591}\refaddr{label1}, W. R\.zysko\orcid{0000-0001-9806-6056}\refaddr{label1}}

\addresses{
\addr{label1} Department of Theoretical Chemistry, Institute of Chemical Sciences, Faculty of Chemistry, Maria Curie-Sk\l odowska University in Lublin, Lublin, Poland
\addr{label2} Departamento de Qu\'{i}mica F\'{i}sica, Facultad de Ciencias Qu\'{i}micas, Universidad Complutense, Madrid 28040, Spain
}


\date{Received 18 November 2025; revised 22 January 2026; accepted 24 January 2026; published 30 March 2026}
\begin{document}

\maketitle

\begin{abstract}
Computer simulations are employed to investigate the adsorption mechanisms of ethane on both homogeneous and inhomogeneous substrates. For homogeneous surfaces, the full range of surface phase transitions --- from incomplete to complete wetting --- can be accessed by tuning the strength of the surface potential. The resulting layering transition temperatures show excellent agreement with experimental measurements of ethane on graphite. By contrast, although all inhomogeneous substrates exhibit a prewetting transition, the adsorption mechanisms are strongly influenced by the stripe width.
\printkeywords  
\end{abstract}

\section{Introduction}

Wetting is an omnipresent phenomenon that can be observed in nature and is widely used for industrial purposes. The capability to control the surface structure permits to tune the propensity to attract or repel a given substance. This manipulation of liquid spreading over the surface offers a great potential for creating  protective coatings and paints \cite{arjmandi24, ejenstam13}, waxes for the skis and skates \cite{kalliorinne25}, deposition of pesticides on plant leaves \cite{bergeron00}, and other applications \cite{bertrand02, bonn09}. Therefore, it seems natural that it attracted a tremendous attention in the theoretical community, aiming at elucidating the microscopic mechanisms underlying this phenomenon \cite{degennes85}. 

The spreading of a liquid over a solid surface is governed by the balance of interfacial tensions, first formulated by Young in 1805, whose celebrated equation provides the basis for determining the contact angles~\cite{young1805}. Later in that century, Gibbs introduced a concept of a surface excess concentration and demonstrated that adsorption changes the interfacial free energy \cite{gibbs1878}. Building on these foundations, the theories for adsorption date back to Freundlich \cite{freundlich07} and Langmuir \cite{langmuir18} in the early twentieth century although these were limited to a single monolayer. This limitation was overcome by the Brunauer-Emmett-Teller theory which extended Langmuir’s approach to describe multilayer adsorption \cite{brunauer38}. More sophisticated theoretical treatments of surface phase transitions were later developed by Cahn \cite{cahn77} and by Ebner and Saam \cite{ebner77}. The phase behavior of multilayer adsorption and the sequence of surface phase transitions driven by increasing the substrate attraction were clarified in the 1980s \cite{pandit82, binder88} and subsequently confirmed experimentally. These studies revealed the key phenomena such as layering~\cite{nham88, kruchten05}, prewetting~\cite{kellay93, taborek92}, and critical wetting transitions \cite{ross01, rafai04}.

Stefan Soko\l owski, to whom the issue is dedicated, has made profound and lasting contributions to the field, many of which are summarized in a comprehensive review article co-authored with his close colleagues Kurt Binder and Andrzej Patrykiejew \cite{patrykiejew00}. Throughout his scientific career, Stefan combined a deep physical insight with remarkable computational skills. He was among the pioneers in Poland to apply computer simulations --- including advanced Monte Carlo and molecular dynamics methods --- to the study of physicochemical processes. {\color{black} One of the principal Stefan's achievements is presented in his book \textit{Statistical Surface Thermodynamics} \cite{patrykiejew16}, which laid foundations for a consistent theoretical description of wetting and surface phase transitions in associating fluids such as water. This theory yields experimentally accessible quantities, most notably the contact angle \cite{dabrowska22, bryk20, terzyk19}, thereby providing a direct link between theory, simulation and experiment.} His curiosity also drove important advances in the development of classical density functional approaches to the investigation of surfaces, slit-like pores, and other confined systems characterized by energetic or structural heterogeneity \cite{karykowski94, chmiel94a, chmiel94b}. 

In this work we employ computer simulations to explore the wetting behavior of ethane on homogeneous and inhomogeneous substrates. The latter systems are essentially the ones proposed by Stefan Soko\l owski over 30 years ago \cite{chmiel94b} and we show that his ideas remain relevant so far. The surfaces considered were structureless and modelled by Lennard-Jones (9,3) potential in which we have modified the attraction strength to encompass the entire range of surface phase transitions from incomplete wetting to complete wetting scenarios. We find that this model successfully reproduces the key features observed in experimental measurements of ethane on graphite \cite{nham88}. By contrast, although all inhomogeneous substrates exhibit a prewetting transition, both the wetting temperature and the adsorption mechanism can be controlled through the stripe width. 

\section{Methods}
\subsection{Force field parameters}
\begin{figure}
    \centering
   \includegraphics[width=0.75\linewidth]{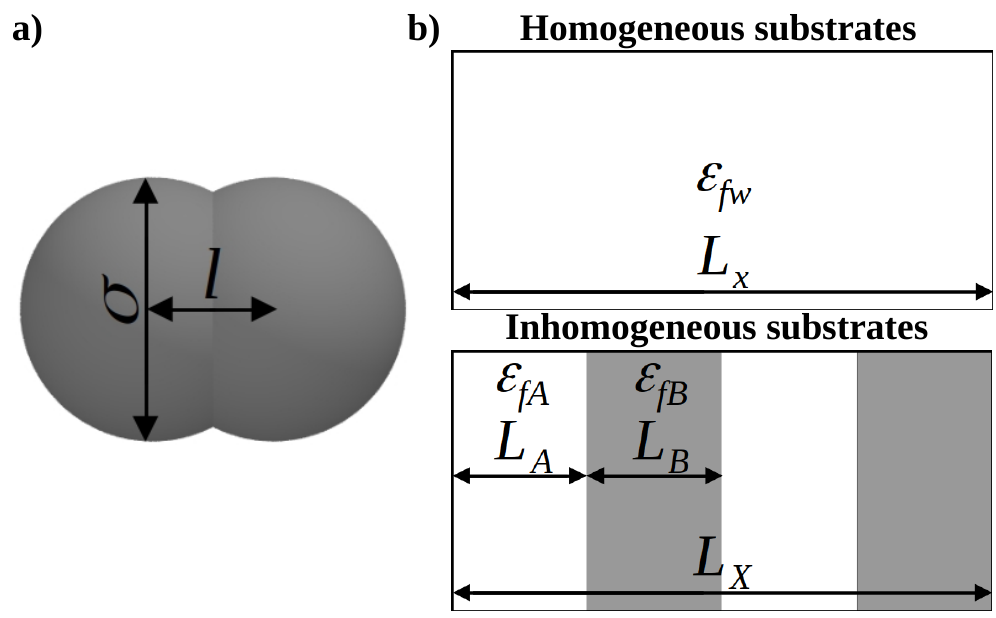}
    \caption{Part (a): Schematic representation of the ethane molecule modelled using the TraPPe force field. Part (b): Sketches of the homogeneous and inhomogeneous substrates. }
    \label{fig:model}
\end{figure}

Ethane was modelled using the TraPPE force field \cite{martin98} which is a well-established model known for accurately describing the phase equilibria of organic molecules. In this representation, the CH$_3$ groups are treated as a single united atom  with a size  of $\sigma_{\text{CH}_3}=0.377$~nm and the distance between the two groups is fixed at $l=0.154$~nm (figure~\ref{fig:model}a). The molecules interact via the standard Lennard-Jones~(12,6) potential with an energy well depth of $\varepsilon_{\text{CH}_3-\text{CH}_3}=98.1$~K, which was set the unit of energy $\varepsilon\equiv\varepsilon_{\text{CH}_3-\text{CH}_3}$. The interactions were truncated at a cutoff distance of $r_{\text{cut}}=3.7\sigma_{\text{CH}_3}\approx1.4$~nm, {\color{black} as suggested in~\cite{martin98}}.  

The interactions between the ethane and substrates were modelled using the Lennard-Jones~(9,3) potential defined as: 

\begin{equation}
    U(z)=\varepsilon_{fw}\left [\frac{2}{15} \left( \frac{\sigma}{z} \right)^9 -
    \left( \frac{\sigma}{z} \right)^3 \right] \quad \text{if} \quad z < 5\sigma,
\end{equation}

\noindent where $\varepsilon_{fw}$ denotes the strength of the fluid-wall interactions, and $\sigma=\sigma_{\text{CH}_3}$ represents the fictitious diameter of the wall atoms. The attractive surface was placed at the bottom of the system, while the reflective wall was positioned at the top. 

We have considered both homogeneous and inhomogeneous substrates. In the former case, only the parameter $\varepsilon_{fw}\in [ 4\varepsilon,6\varepsilon]$ was varied to investigate different wetting scenarios. For the inhomogeneous surfaces, the substrate consisted of alternating stripes of equal width $L_A=L_B=L_S$, as shown in figure~\ref{fig:model}b. The width of the stripes  was varied as $L_S=(2.0\sigma_{\text{CH}_3}, 6.0\sigma_{\text{CH}_3}, 12.0\sigma_{\text{CH}_3})$, meaning that the number of stripes decreases as their size is increased. The interactions strengths were fixed at $\varepsilon_{fA}=4\varepsilon$ and $\varepsilon_{fB}=6\varepsilon$. 

The system sizes used in the simulations for both the bulk and surface configurations are listed in table~\ref{tab:sys-size}. In the former, periodic boundary conditions were applied in all three directions, while in the latter, only in the lateral $x$, $y$ directions.

\begin{table}[h!]
    \centering
    \caption{System sizes for different systems studied.}
    \vspace{2mm}
    \begin{tabular}{|cc|}
    \hline
        Systems & $L_x \times L_y \times L_z$ (nm$^3$)\\\hline
        \multirow{3}{*}{Bulk} & $12\sigma_{\text{CH}_3}\times 12\sigma_{\text{CH}_3} \times12\sigma_{\text{CH}_3}$  \\
                              & $16\sigma_{\text{CH}_3}\times 16\sigma_{\text{CH}_3} \times16\sigma_{\text{CH}_3}$ \\
                              & $18\sigma_{\text{CH}_3}\times 18\sigma_{\text{CH}_3} \times18\sigma_{\text{CH}_3}$ \\ \hline
        Homogeneous substrates & $14\sigma_{\text{CH}_3}\times 14\sigma_{\text{CH}_3}\times30\sigma_{\text{CH}_3}$ \\\hline
        Inhomogeneous substrates& $24\sigma_{\text{CH}_3} \times14\sigma_{\text{CH}_3}\times30\sigma_{\text{CH}_3}$\\\hline
    \end{tabular}
    \label{tab:sys-size}
\end{table}

\subsection{Simulation protocol}

We conducted two distinct sets of simulations. To determine the isotherms for different substrates, regular Monte Carlo simulations were performed in the grand canonical ensemble.  Each simulation included an equilibration period of $10^6$ Monte Carlo steps, followed by production runs of $10^7$ steps for data collection. 

To evaluate the phase diagrams, the hyper-paralleling tempering method proposed by Yan and de Pablo~\cite{yan99} with histogram reweighting technique \cite{ferrenberg88} was employed. To determine the coexistence lines, 30 replicas were used for the bulk systems, while under confinement, the number varied between 10 and 15, depending on the type of the surface phase transition. These simulations required considerably higher statistics, with the averages collected over $10^9$ Monte Carlo steps after an initial equilibration for $10^6$ steps. 

A single Monte Carlo step consisted of the attempts to (i) translate and rotate a randomly selected molecule, (ii) insert a new molecule or (iii) remove one from the system. 

\subsection{Properties}
\subsubsection{Critical temperature}\label{sec:critical}

The critical temperature, $T_C$, was estimated using the standard procedure proposed by Binder \cite{binder81}, in which the fourth-order cumulant is calculated, defined as

\begin{equation}
    U_L=1-\frac{\langle m^4\rangle}{3\langle m^2\rangle^2}\,,
\end{equation}
where the order parameter is defined as $m=\rho-\langle\rho\rangle$ \cite{rzysko12}. {\color{black} $U_L$ is then estimated for different system sizes $L$ as a function of temperature and the critical temperature, was then estimated as the intersection point of these curves.}

\subsubsection{Coexistence chemical potential}\label{sec:coex}

Coexistence chemical potential, required for the proper analysis of surface phase transitions, was estimated using two distinct methods. The first involved reweighting the two-dimensional histograms of the system's energy $U$ and the number of molecules $N$, denoted as $P(N,U)$. For sufficiently long simulation runs at various chemical potentials $\mu$ and temperatures $T$, the system could be found in either of two phases, which is manifested by two well-separated peaks in the histogram. For a given temperature, the chemical potential was varied until the areas under the two peaks became equal, corresponding to the equal free energies of the two phases --- and thus to the coexistence chemical potential.

In the second method, simulations were performed in the grand canonical ensemble at an arbitrary chemical potential $\mu_0$ but the number of molecules in the system was allowed to fluctuate only by the amount of $\pm5$ about the initial value. This constraint was implemented via an additional weighting function \cite{benet14, macdowell18}. During these simulations, histograms of the number of molecules in the system $P_0(N)$, were collected. From this distribution, the corresponding probability at the coexistence chemical potential $\mu_C$ could be estimated as

\begin{equation}
    P_C(N)=P_0(N) \re^{\beta(\mu_C-\mu_0)N},
    \label{eq:luis}
\end{equation}

\noindent with $\beta=1/k_\text{B}T$. Noting that $\ln P_C(N)$ should be a linear function, the slope provides the difference between the coexistence and current chemical potentials, i.e., $\beta(\mu_C-\mu_0)$. 

\section{Results and discussion}
\subsection{Liquid-vapor phase equilibria}

\begin{figure} [h!]
    \centering
    \includegraphics[width=\linewidth]{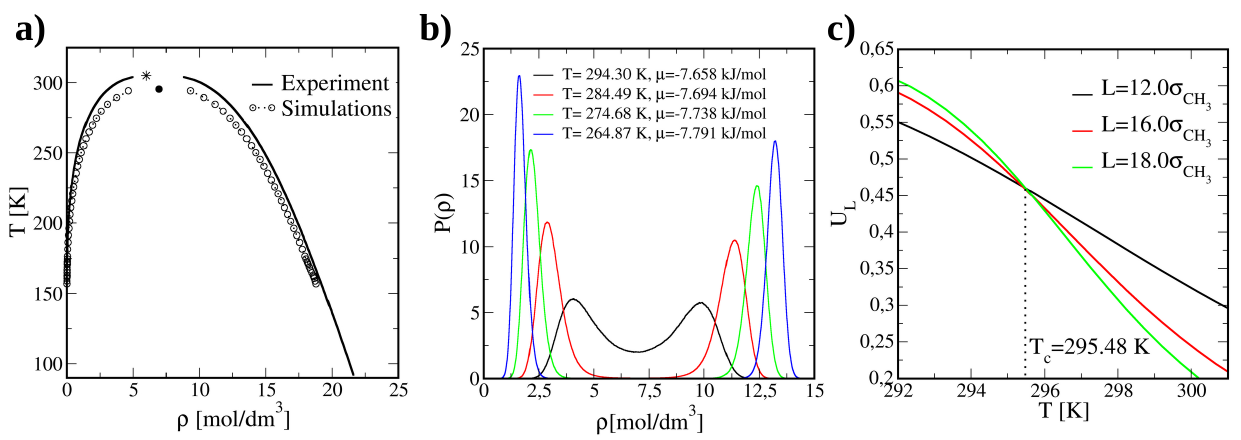}
    \caption{(Colour online) Part (a): Bulk phase diagram in the temperature-density plane, with critical points indicated by an asterisk for experiments and a filled circle for simulations. Part (b): Examples of probability distributions $P(\rho)$ obtained from the hyper-parallel tempering method. Part (c): Temperature dependence of the Binder's cumulant for three system sizes. }
    \label{fig:diag}
\end{figure}

To describe the surface phase transitions, the reference phase diagram was first evaluated.
Figure~\ref{fig:diag}a shows the liquid-vapor coexistence lines up to the critical point, which are in good agreement with experimental results \cite{friend91}. Discrepancies become noticeable only at higher temperatures. Several histograms obtained from the hyper-parallel tempering at different temperatures are shown in figure~\ref{fig:diag}b, illustrating the coexistence densities of the two phases. Additionally, the coexistence chemical potentials are indicated in the legend of this figure. 

It is well-known that approaching the critical point leads to large density fluctuations and critical slowing down of the dynamics, making its accurate evaluation challenging, even when using hyper-parallel tempering. To overcome these difficulties, the finite-size scaling method proposed by Binder as described in section~\ref{sec:critical} can be used. Binder's cumulants were calculated along the liquid-vapor coexistence lines for three system sizes $L=12\sigma_{\text{CH}_3},16\sigma_{\text{CH}_3},18\sigma_{\text{CH}_3}$ as shown in figure~\ref{fig:diag}c. The intersection point of these three curves represents a fixed point, yielding a critical temperature $T_C=295.48$~K, a bit below the experimental value, $T_{C,\textrm{exp}}=305.33$~K \cite{friend91}. 

\subsection{Homogeneous substrates}

\begin{figure}[h!]
    \centering
    \includegraphics[width=0.75\linewidth]{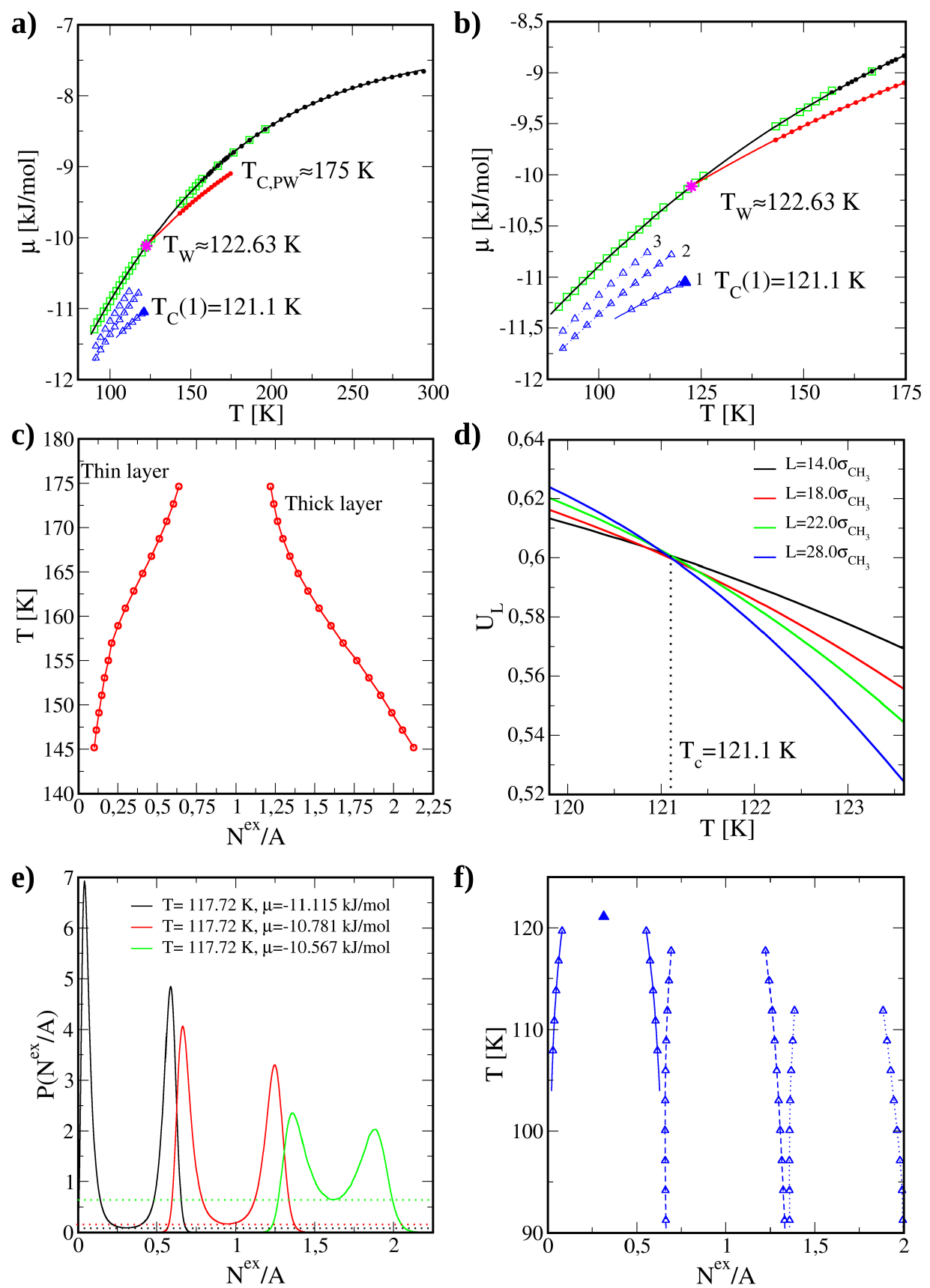}
    \caption{(Colour online) Part (a): Bulk (black and green points) and surface (red and blue points) phase diagrams in the $\mu-T$ plane. Green square points are estimated using equation~\ref{eq:luis} while the remaining ones are obtained from the equal area rule as described in section~\ref{sec:coex}. 
    Red circles are the results for the substrate with $\varepsilon_{fw}=5\varepsilon$ where the prewetting transition is observed, starting at $T_W$ marked as a pink asterisk, and terminating in the critical prewetting point, T$_{C,PW}$. 
    Blue triangles correspond to the layering transitions occurring in the system with $\varepsilon_{fw}=6\varepsilon$. Magnified region emphasizing the layering transitions is shown in part (b). 
    Part (c): Phase diagram of the thin-thick film transition for $\varepsilon_{fw}=5\varepsilon$. Part (d): Temperature dependence of the Binder's cumulant for four system sizes for the first layering transition. 
    Part (e): Probability distributions of excess adsorption per unit area for systems with different number of layers. Part (f): Phase diagram of the layering transitions for  $\varepsilon_{fw}=6\varepsilon$. Critical point for the first layering is marked as a filled triangle.}
    \label{fig:homo}
\end{figure}

Before discussing the surface behavior of different substrates, the bulk phase diagram in the $\mu-T$ plane (figure~\ref{fig:homo}a, b) was evaluated. Both methods described in section~\ref{sec:coex} yield quantitatively identical results. The figure also shows the results for homogeneous substrates. For the weakest adsorbing system, $\varepsilon_{fw} = 4\varepsilon$, no wetting transition could be identified up to the critical point. Increasing the adhesion strength to $\varepsilon_{fw} = 5\varepsilon$ resulted in the wetting transition at $T_W=122.63$~K, estimated as the intersection of the bulk and surface coexistence lines. Above $T_W$, a thin–thick film transition occurs as the chemical potential approaches the bulk coexistence value. This is evident in figure~\ref{fig:homo}c, which shows a sharp increase in the excess adsorption per unit area, $N^{\text{ex}}/A$, with $N^{\text{ex}}$ defined as:

\begin{equation}
    N^{\text{ex}}(\mu)=\langle N(\mu)\rangle-\rho_bV \re^{\beta(\mu_0-\mu_C)}.
\end{equation}

The prewetting (thin-thick film) transition is observed for temperatures up to the critical prewetting point, estimated at $T_{C,PW}\approx175$~K. In the temperature range $T_W<T<T_{C,PW}$ a complete wetting occurs above the prewetting line. Moreover, at temperatures above $T_{C,PW}$, complete wetting is observed at any chemical potential approaching the bulk coexistence line.

For the substrate with $\varepsilon_{fw}=6\varepsilon$,  complete wetting is already observed from the triple point, $T_{t}=90.352$~K. In this case, a series of the layering transitions occurs, as shown in figure~\ref{fig:homo}a, b. Each transition exhibits a distinct critical layering temperature, $T_C(n)$. In order to estimate their values, the Binder's cumulant for several system sizes was evaluated for the first layering transition. This is shown figure~\ref{fig:homo}d, and the critical temperature was estimated to be $T_C(1)=121.1$~K (marked as a filled triangle in figure~\ref{fig:homo}f). The critical temperatures of the next two layering transitions were obtained by comparing the probability distribution functions of the excess adsorption per unit surface, displayed in figure~\ref{fig:homo}e. It is evident that the minima of these distributions-corresponding to the line tension --- increase at a fixed temperature as the number of adsorbed layers grows, as indicated by the dotted lines in figure~\ref{fig:homo}e.  This increase indicates that the system approaches the critical temperature, where the surface tension vanishes. Thus, the critical temperature of the following layering transitions decreases with increasing number of layers $n$, which is schematically shown in the phase diagram in figure~\ref{fig:homo}f. 

This trend contrasts with the results for lattice systems \cite{pandit82} and some experimental measurements \cite{ramesh84}. However, for alkanes adsorbed on graphite --- experimental systems closely related to the current model --- an identical trend is observed: $T_C(n)$ decreases with increasing $n$ \cite{kruchten05, nham88}. Moreover, our estimates of $T_C(n)$ are in a very close quantitative agreement with experimental results for ethane adsorbed on graphite \cite{nham88}, as summarized in table~\ref{tab:Tc_comparison}. The small discrepancies are mainly due to a slight shift in the bulk phase diagram (cf. figure~\ref{fig:diag}a) between the TraPPe model and experimental data. It should be noted that the first layering transition occurs at the pressure (chemical potential) ranges too small to be accessed experimentally, thus experimental data start from the $T_C(2)$.

\begin{table}[h!]
    \centering
    \caption{Comparison of critical layering temperatures $T_C(n)$ estimated from Monte Carlo simulations with experimental data taken from~\cite{nham88}.}
    \vspace{2mm}
    \begin{tabular}{|c|c|c|}
    \hline
    \multicolumn{3}{|c|}{$T_C(n)$ [K]}     \\\hline
      n & Experiment & Simulations \\\hline
      1 & - & $121.1\pm0.4$ \\
      2 & $120.8\pm0.3$ & $119.5\pm1.5$ \\
      3 & $116.1\pm0.6$ & $115.7\pm1.5$ \\\hline
    \end{tabular}
    \label{tab:Tc_comparison}
\end{table}

\subsection{Inhomogeneous substrates}

\begin{figure}[h!]
    \centering
    \includegraphics[width=0.75\linewidth]{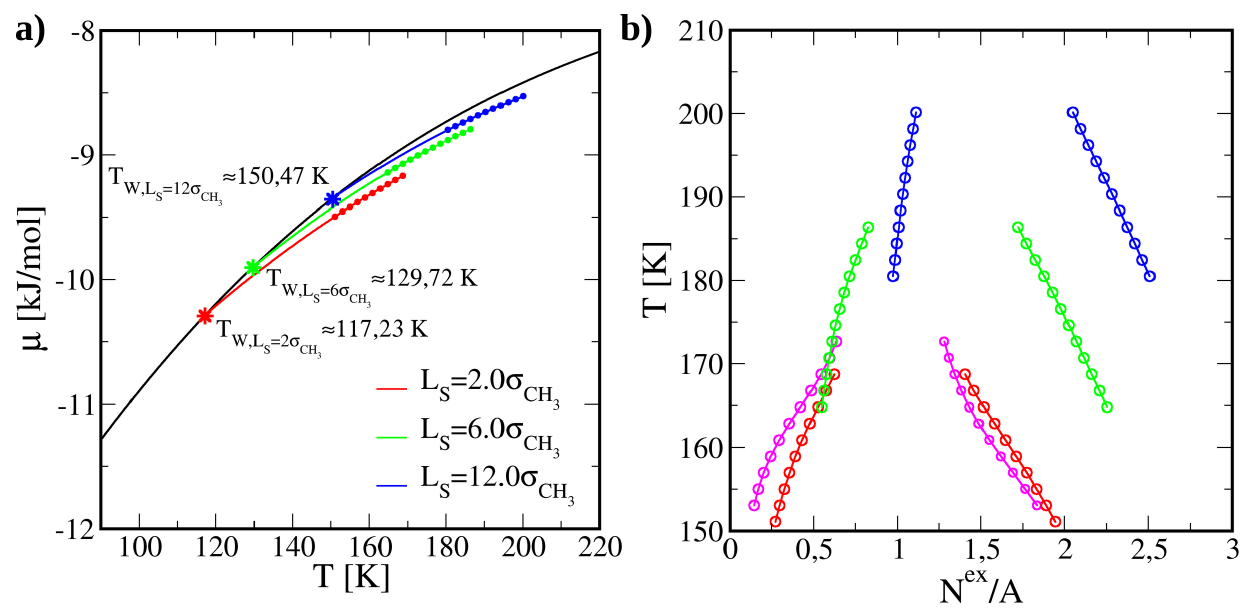}
    \caption{(Colour online) Part (a): Bulk (black solid line) and surface (red, green and blue points) phase diagrams in the $\mu-T$ plane. Part (b): Phase diagram of the thin-thick film transition for different stripe widths and homogeneous substrate $\varepsilon_{fw}=5.0$ (magenta circles). }
    \label{fig:inhomo}
\end{figure}

Knowing the wetting behavior of homogeneous surfaces, it is possible to assess how the inhomogeneous substrates comprised of alternating stripes modify the surface phase transitions.  Since these substrates are comprised of low- ($\varepsilon_{fw}=4\varepsilon$) and high-energy ($\varepsilon_{fw}=6\varepsilon$) regions, two limiting scenarios can be aniticipated: (i) for infinitely narrow stripes, the substrate field should effectively average out to that of a homogeneous substrate with $\varepsilon_{fw}=5\varepsilon$; and (ii) for infinitely wide stripes, the two regions should behave as two independent surfaces. To examine these assumptions, three stripe widths were considered. The resulting phase diagram is shown in figure~\ref{fig:inhomo}.

\begin{figure}[h!]
    \centering
    \includegraphics[width=0.75\linewidth]{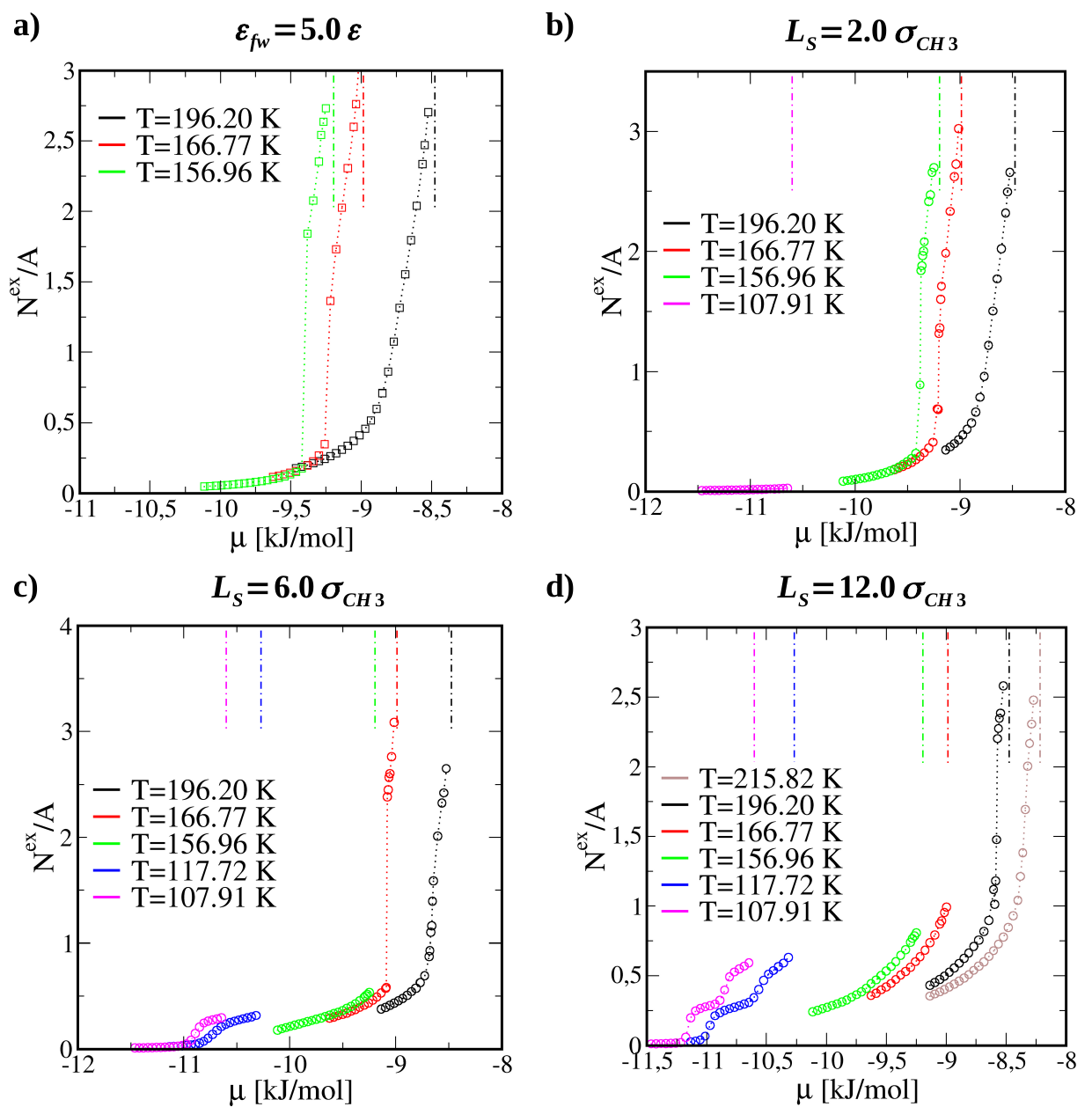}
    \caption{(Colour online) Part (a): Excess adsorption isotherms  $N^{\text{ex}}/A$ as a function of chemical potential~$\mu$ calculated for a homogeneous substrate with $\varepsilon_{fw}=5.0\varepsilon$ and (b-d) three different stripes width $L_S$. Dash-dotted lines indicate the bulk coexistence chemical potential $\mu_C$ at the corresponding temperature.}
    \label{fig:inhomo-ads}
\end{figure}

Overall, the wetting scenario closely resembles that of the homogeneous surface with $\varepsilon_{fw}=5\varepsilon$: in all cases, a prewetting line associated with the thin–thick film transition appears above $T_W$. The wetting transition temperature increases with stripe width, except for the narrowest stripes of $L_S=2\sigma_{\text{CH}_3}$, for which $T_W$ is nearly identical to the homogeneous $\varepsilon_{fw}=5\varepsilon$ case. This behavior may arise either from uncertainties in determining the phase diagram or from an additional potential source generated at the stripe boundaries by adsorbed particles. The latter is due to the stripe width in this smallest case being shorter than the range of the interparticle potential, allowing particles to interact simultaneously with both low- and high-energy regions. Consequently, even a narrower stripe width needs to be considered for the potential to be averaged out into the homogeneous case of $\varepsilon_{fw}=5\varepsilon$. For wider stripes, we expect them to behave more like independent surfaces, thereby shifting the transition temperatures to higher values, which is readily observed in figure~\ref{fig:inhomo}.

As a natural consequence to the shift in $T_W$, both the prewetting line and critical prewetting temperature move into higher values in comparison with the homogeneous surfaces. This behavior is illustrated in figure~\ref{fig:inhomo-ads} which shows adsorption isotherms at several temperatures chosen to lie below  $T_W$, above $T_W$, and above $T_{C,PW}$. It is clear that the adsorption isotherms for homogeneous and the narrowest stripe widths are almost identical (cf. figure~\ref{fig:inhomo-ads}a,b). By contrast, wider stripes shift the temperature scale to higher values, although the adsorption mechanism remains unchanged. All of these observations are fully consistent with the phase diagrams shown in figure~\ref{fig:inhomo}. 

\begin{figure}[h!]
    \centering
    \includegraphics[width=0.75\linewidth]{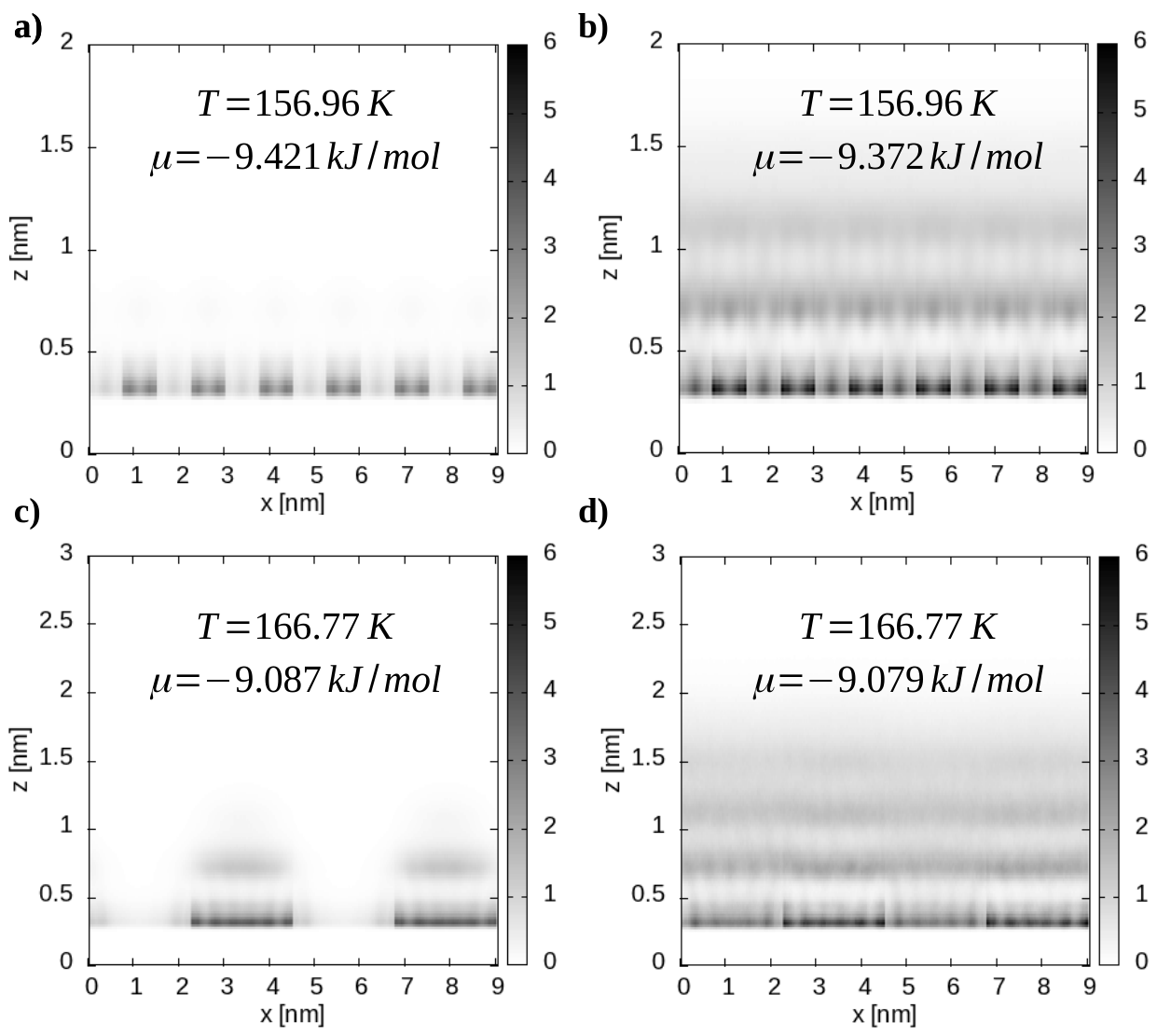}
    \caption{Two-dimensional density profiles $\rho(x,z)$ for stripe widths $L_S=2.0\sigma_{\text{CH}_3}$ (a, b) and $L_S=6.0\sigma_{\text{CH}_3}$ (c, d).
    } 
    \label{fig:2d-part1}
\end{figure}

However, it needs to be noted that the excess adsorption shown in this figure is calculated for the entire system which do not provide any information on how adsorption proceeds. One needs to bear in mind that for wider stripes there are large low- and high-energy regions that, on their own, exhibit incomplete and complete wetting, respectively (cf. figure~\ref{fig:homo}a). Therefore, the evolution of the film growth as the bulk coexistence potential is approached needs to be carefully examined. This is illustrated in figure~\ref{fig:2d-part1}. 

For the narrowest stripes, it is readily seen that the highest adsorption density can be found in the high-energy stripes, although a certain amount of ethane molecules is also adsorbed in the the low-energy regions (figure~\ref{fig:2d-part1}a). By contrast, for the intermediate stripe width ($L_S=6\sigma_{\text{CH}_3}$) the adsorption is mainly observed in the high-energy stripes, although there is a noticeable number of molecules accumulated near the boundaries. This behavior arises because the droplets formed on the high-energy regions must adjust their shape to the free energy minima which requires spreading beyond the stripe edges and forming a finite contact angle (figure~\ref{fig:2d-part1}c). Despite these geometric differences, the mechanism of film formation beyond the thin–thick transition remains the same for both stripe widths, as shown in figures~\ref{fig:2d-part1}b and \ref{fig:2d-part1}d.

\begin{figure}[h!]
    \centering
    \includegraphics[width=\linewidth]{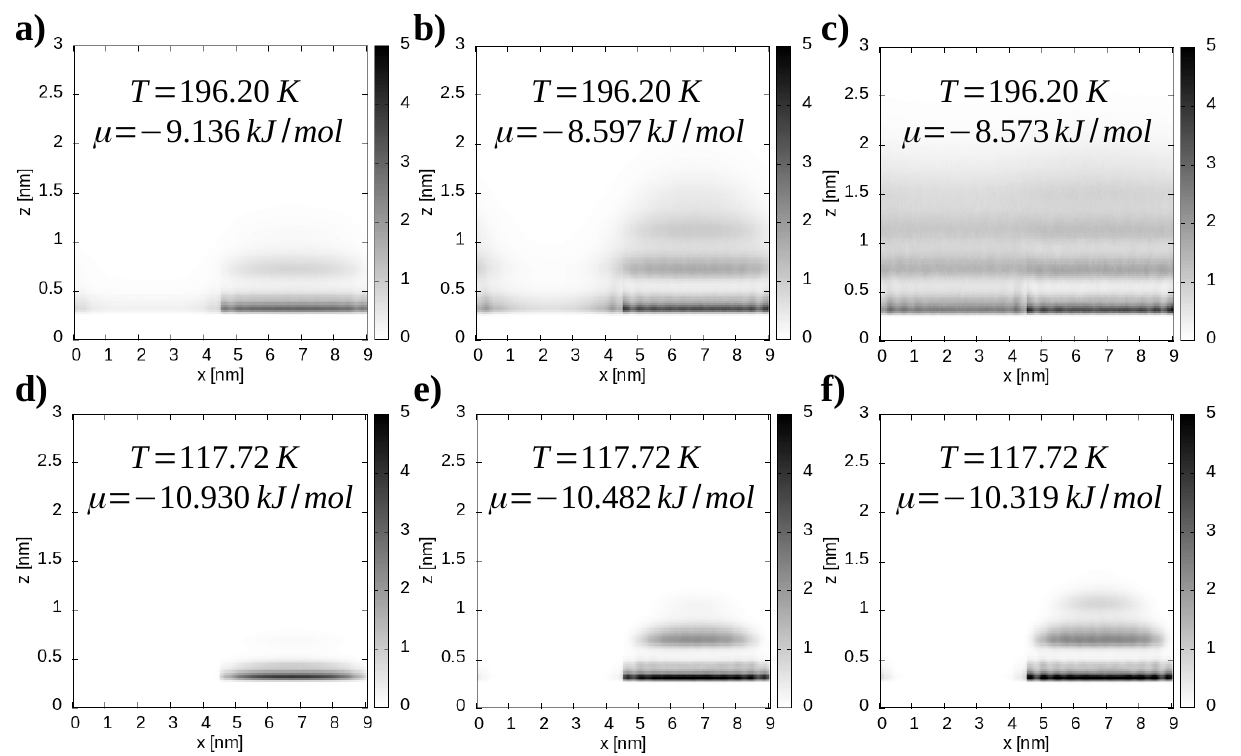}
    \caption{Two-dimensional density profiles $\rho(x,z)$ for a stripe width $L_S=12.0\sigma_{\text{CH}_3}$.} 
    \label{fig:2d-part2}
\end{figure}

Adsorption on the high-energy stripes becomes even more pronounced for the widest stripes $L_S=12\sigma_{\text{CH}_3}$, as shown in figure~\ref{fig:2d-part2}a. Nevertheless, an identical thin-thick film transition observed for narrower stripes still occurs, although the corresponding temperature scale is shifted towards higher values. What is more interesting is the surface behavior at low temperatures, considerably below the wetting temperature $T_W$. The system exhibits a pseudo-layering transition, reminiscence of the behavior seen on homogeneous substrates with $\varepsilon_{fw}=6\varepsilon$, hints of which were already visible in figure~\ref{fig:inhomo-ads}d. This is illustrated in figure~\ref{fig:2d-part2}d--f. It can be anticipated that for sufficiently wide stripes, the true layering transitions should emerge, analogous to those observed on homogeneous surfaces (cf. figure~\ref{fig:homo}a). Achieving this limit would require stripes large enough to behave effectively as independent substrates. This interpretation is further supported by the fact that similar, although much weaker, signatures of layering are already observed for the intermediate stripe width (cf. figure~\ref{fig:inhomo-ads}c).

\section{Conclusions}

In this paper, it was demonstrated that the TraPPE potential can be successfully employed to investigate the wetting behavior of ethane on various substrates. By tuning the strength of the external surface potential for homogeneous surfaces, the full range of surface phase transitions was accessed --- including incomplete wetting, prewetting, and layering. Moreover, in the case of layering transitions, the model accurately reproduces the critical layering temperatures experimentally measured for ethane adsorbed on graphite, proving the usefulness of the model system studied. 

The situation becomes more complex when inhomogeneous substrates are considered. For all examined stripe widths, a prewetting transition is observed. However, the wetting transition temperature increases with stripe width, except for the narrowest stripes of $2\sigma_{\text{CH}_3}$. Consequently, the same trend is found for the corresponding critical prewetting temperatures. 

Although each inhomogeneous substrate exhibits the same surface phase transition, the difference lies in how the particles are adsorbed on the surface. For the narrowest stripes, the adsorbed film is distributed almost uniformly across the entire surface, effectively approximating the homogeneous substrate $\varepsilon_{fw}=5\varepsilon$. To observe this effect even more clearly, stripes of smaller width would need to be examined. By contrast, for stripe widths $L_S=6\sigma_{\text{CH}_3}$ and $L_S=12\sigma_{\text{CH}_3}$, droplets with finite contact angles form in the high-energy regions. Moreover, for $L_S=12\sigma_{\text{CH}_3}$, the reminiscence of layering transitions is observed at $T=117.72$~K --- well-below the wetting temperature $T_W$. This suggests that for sufficiently wide stripes, the individual regions should begin to behave as independent, homogeneous surfaces. 

We believe that these findings provide new insights into how nanoscale surface heterogeneity controls the wetting behavior and may guide the design of tailored interfaces for adsorption and coating applications.

\section{Acknowledgments}
We would like to dedicate this manuscript to our dearest colleague, Stefan Soko\l owski who sadly passed away on 24 June 2024. Stefan was an extraordinarily influential person, not only in shaping our scientific understanding but also in inspiring all of us personally and professionally. His legacy endures in the pioneering work he carried out in introducing and advancing computer simulations in Poland, in the people he inspired, and in the countless discussions, collaborations, and friendships he fostered throughout his remarkable career.
We would also like to thank Luis G. MacDowell for invaluable discussions.

\bibliography{biblio}
\bibliographystyle{cmpj}

\ukrainianpart

\title{Як етан змочує різні підкладки?}
\author{Л. Баран, \refaddr{label1,label2}
	Д. Тарасевич\refaddr{label1}, В. Жисько \refaddr{label1}}

\addresses{
	\addr{label1}
	Кафедра теоретичної хімії, Інститут хімічних наук, Хімічний факультет, Університет Марії Склодовської-Кюрі, 20031 Люблін, Польща
	\addr{label2} Кафедра фізичної хімії, факультет хімічних наук, Університет Комплутенсе, Мадрид 28040, Іспанія
}

\makeukrtitle

\begin{abstract}
	\tolerance=3000%
	Для дослідження механізмів адсорбції етану як на однорідних, так і на неоднорідних підкладках
	використовується метод комп'ютерного моделювання. Для однорідних поверхонь повний діапазон фазових переходів поверхні --- від неповного до повного змочування --- можна отримати, змінюючи силу поверхневого потенціалу. Отримані температури переходу нашарування демонструють чудову відповідність з експериментальними вимірюваннями для етану на графіті. На противагу цьому, хоча всі неоднорідні підкладки демонструють перехід попереднього змочування, механізми адсорбції сильно залежать від ширини смуги.

	\keywords поверхневий фазовий перехід, комп'ютерне моделювання, адсорбція
\end{abstract}
\lastpage
\end{document}